\documentstyle[aps,epsf,prl]{revtex}
\begin{document}
\twocolumn[ 
\phantom{a}
\vskip0.2in
\preprint{\vbox{\hbox{UCSD/PTH 95--12} \hbox{CALT--68--2012}
\hbox{hep-ph/9508210} }}
\title{Chiral Perturbation Theory for $\phi\rightarrow \rho \gamma
\gamma$ and $\phi \rightarrow \omega \gamma \gamma$}
\author{Adam K. Leibovich${}^a$, Aneesh V. Manohar${}^b$, and Mark B. Wise
${}^a$}
\address{${}^a$\ California Institute of Technology, Pasadena, CA 91125}
\address{${}^b$\ Department of Physics,
University of California at San Diego\\
9500 Gilman Drive, La Jolla, CA 92093-0319}
\date{July 1995} \maketitle \widetext
\vskip-2.0in
\rightline{\vbox{ \hbox{UCSD/PTH 95--12} \hbox{CALT--68--2012}
\hbox{hep-ph/9508210} }}
\vskip1.5in
\begin{abstract}
We predict differential decay distributions for $\phi\rightarrow \rho \gamma
\gamma$ and $\phi \rightarrow \omega \gamma \gamma$ using chiral perturbation
theory. We also consider the isospin violating decay $\phi \rightarrow \omega
\pi^0$. Experimental information on these decays can be used to determine
couplings in the heavy vector meson chiral Lagrangian.

{\bf ERRATUM: It was shown by P. Ko et al., Phys. Lett. B366 (1996) 287, that 
there is a $\eta^\prime$ pole contribution that dominates over what
we calculated.}
\end{abstract}


] 

\narrowtext

The $\phi$ factory being built at Frascati is expected to produce $10^{11}$
$\phi$'s per year~\cite{DAPHNE}, enabling rare decay modes of the $\phi$ to be
measured for the first time. Some of these rare decays are of the form $V
\rightarrow V^\prime X$, where $V$ and $V^\prime$ are vector mesons, and can be
studied using chiral perturbation theory because the $V-V^\prime$ mass
difference is small. The chiral Lagrangian for $V \rightarrow V^\prime X$
transitions written in terms of heavy meson fields was given in a previous
paper~\cite{jmw}. This form of the chiral Lagrangian for matter fields has a
well-defined momentum expansion~\cite{jm}, and can be used as the starting
point for a systematic expansion in powers of momentum.  The vector meson
chiral Lagrangian has coupling constants (analogous to $g_A$) which have not
been measured. In this paper, we show that one can determine the unknown
coupling constants using the rare decay modes $\phi\rightarrow \omega \gamma
\gamma$ and $\phi \rightarrow \rho \gamma \gamma$.

The vector meson fields can be written as a $3 \times 3$ octet
matrix\footnote{We will use the same notation as Ref.~\cite{jmw}. The notation
for the pseudoscalar meson fields is standard, and will not be repeated
here. The pion decay constant is $f=132$~MeV.}
\begin{equation}\label{eq:1}
{\cal O}_\mu = \left[
\begin{array}{ccc}
{\rho^0_\mu \over \sqrt{2}} + {\phi_\mu^{(8)} \over \sqrt{6}} & \rho_\mu^+ &
K_\mu^{*+} \\ \rho_\mu^- & -{\rho_\mu^0 \over \sqrt{2}} + {\phi_\mu^{(8)} \over
\sqrt{6}} & K_\mu^{*0} \\ K_\mu^{*-} & \overline K_\mu^{*0} & -{2\phi_\mu^{(8)}
\over \sqrt{6}}
\end{array} \right],
\end{equation}
and as a singlet
\begin{equation}
S_\mu = \phi_\mu^{(0)}. \label{eq:2}
\end{equation}
The $\phi$ and $\omega$ mass eigenstates are linear combinations of
$\phi^{(0)}$ and $\phi^{(8)}$,
\begin{mathletters}
\label{eq:5}
\begin{eqnarray}
|\phi\rangle &=& \sin \Theta_V|\phi^{(0)} \rangle - \cos \Theta_V | \phi^{(8)}
\rangle, \\ |\omega \rangle &=& \cos \Theta_V|\phi^{(0)} \rangle + \sin
\Theta_V | \phi^{(8)} \rangle.
\end{eqnarray}
\end{mathletters}

The chiral Lagrange density which describes the interactions of the vector
mesons with the low-momentum $\pi$, $K$ and $\eta$ mesons has the general
structure\footnote{Here we neglect vector meson widths. These can be taken into
account by including antihermitian terms in $\cal L$.}
\begin{equation}\label{eq:3}
{\cal L} = {\cal L}_{\rm kin} + {\cal L}_{\rm int} + {\cal L}_{\rm mass}.
\end{equation}
At leading order in the derivative and quark mass expansions,
\begin{eqnarray}
{\cal L}_{\rm kin} &=& - i\, S_\mu^\dagger (v \cdot \partial) S^\mu - i\, {\rm
Tr}\, {\cal O}_\mu^\dagger (v \cdot {\cal D}) {\cal O}^\mu,\label{kin}
\end{eqnarray}
and
\begin{eqnarray}
{\cal L}_{\rm int} &=& i g_1 \, S_\mu^\dagger\, {\rm Tr}\, \left({\cal O}_\nu
A_\lambda \right) v_\sigma \epsilon^{\mu \nu \lambda \sigma} + h.c. \nonumber\\
&&+ i g_2 \,{\rm Tr} \left( \left\{ {\cal O}_\mu^\dagger, {\cal O}_\nu\right\}
A_\lambda \right) v_\sigma \epsilon^{\mu \nu \lambda \sigma},
\label{eq:4}
\end{eqnarray}
\begin{eqnarray}
{\cal L}_{\rm mass} &=& \mu_0\, S_\mu^\dagger S^\mu + \mu_8\, {\rm Tr}\, {\cal
O}_\mu^\dagger {\cal O}^\mu \nonumber\\ &&+ \lambda_1\, \left({\rm Tr}\, \left(
{\cal O}_\mu^\dagger {\cal M}_\xi \right) S^\mu + h.c \right) \nonumber\\ &&+
\lambda_2\, {\rm Tr}\, \left( \left\{ {\cal O}_\mu^\dagger, {\cal O}^\mu
\right\} {\cal M}_\xi \right)
\label{eq:13} \\
&&+ \sigma_0\, {\rm Tr}\, {\cal M}_\xi \ S_\mu^\dagger S^\mu + \sigma_8\, {\rm
Tr}\, {\cal M}_\xi \ {\rm Tr}\, {\cal O}_\mu^\dagger {\cal O}^\mu , \nonumber
\end{eqnarray}
where $v^\mu$ is the vector meson four-velocity,
\begin{equation}
{\cal D}^\nu {\cal O}^\mu = \partial^\nu {\cal O}^\mu + \left[ V^\nu, {\cal
O}^\mu \right],\label{cov}
\end{equation}
\begin{equation}V^\mu = \frac 1 2
\left(\xi \partial^\mu \xi^\dagger +\xi^\dagger \partial^\mu \xi \right) ,\
A^\mu = \frac i 2 \left(\xi \partial^\mu \xi^\dagger -\xi^\dagger \partial^\mu
\xi \right) ,
\end{equation}
$\cal M$ is the quark mass matrix $ {\cal M} = {\rm diag} \left(m_u, m_d,
m_s\right)$, and
\begin{equation}
{\cal M}_\xi = {1 \over 2} \left( \xi {\cal M} \xi + \xi^\dagger {\cal M}
\xi^\dagger\right).
\end{equation}
The terms in ${\cal L}_{\rm kin}$ appear with minus signs because the
polarization vector is spacelike.

In the large-$N_c$ limit, the octet and singlet mesons can be combined
into a single ``nonet'' matrix~\cite{thooft,ven}
\begin{equation}\label{eq:21}
N_\mu = {\cal O}_\mu + {I\over\sqrt{3}}\, S_\mu,
\end{equation}
which enters the chiral Lagrangian. The kinetic, interaction and mass terms at
leading order in $1/N_c$ are
\begin{eqnarray}
{\cal L}_{\rm kin} &\rightarrow& - i \, {\rm Tr}\, N_\mu^\dagger \left( v \cdot
{\cal D} \right) N^\mu ,
\label{eq:22}
\end{eqnarray}
\begin{equation}\label{eq:22b}
{\cal L}_{\rm int} \rightarrow i g_2\, {\rm Tr}\, \left( \left\{N_\mu^\dagger,
N_\nu\right\} A_\lambda\right) v_\sigma \epsilon^{\mu\nu\lambda\sigma},
\end{equation}
and
\begin{equation}\label{eq:22c}
{\cal L}_{\rm mass} \rightarrow \mu\, {\rm Tr}\, N_\mu^\dagger N^\mu +
\lambda_2 \, {\rm Tr}\,\left( \left\{N_\mu^\dagger, N^\mu\right\} {\cal M}_\xi
\right) .
\end{equation}
Comparing with Eqs.~(\ref{kin})--(\ref{eq:13}), one finds that in the $N_c
\rightarrow \infty$ limit,
\begin{equation}\label{eq:15}
\Delta \mu \rightarrow 0, \ \sigma_0 \rightarrow {2 \lambda_2 \over 3}, \
\sigma_8 \rightarrow 0,
\end{equation}
\begin{equation}\label{eq:20}
g_1 \rightarrow {{2g_2}\over \sqrt{3}},\qquad \lambda_1 \rightarrow {{2
\lambda_2} \over \sqrt{3}},\qquad \tan \Theta_V \rightarrow {1 \over \sqrt2},
\end{equation}
the $|\phi \rangle$ state becomes ``pure'' $|s\bar s \rangle$, and the nonet
matrix is
\begin{equation}\label{eq:23}
N_\mu = \left[
\begin{array}{ccc}
{\rho_\mu^0 \over \sqrt{2}} + {\omega_\mu \over \sqrt{2}} & \rho_\mu^+
&K_\mu^{*+}\\ \rho_\mu^- & -{\rho_\mu^0 \over \sqrt{2}} + {\omega_\mu \over
\sqrt{2}} & K_\mu^{*0}\\ K_\mu^{*-} & \overline K_\mu^{*0} & \phi_\mu
\end{array}
\right]\, .
\end{equation}

In studying $\phi$ decays, it is important to determine the mixing angle
$\Theta_V$ and the couplings $g_1$ and $g_2$.  The mixing angle $\Theta_V$ can
be determined to first order in flavor $SU(3)$ symmetry breaking from the
measured vector meson masses~\cite{Kok},
\begin{equation}\label{eq:19}
\tan \Theta_V = \sqrt{{m_\phi - {4\over3}m_{K^{*}} + {1 \over 3}m_\rho} \over
{{4 \over 3}m_{K^{*}} - {1 \over 3} m_\rho - m_\omega}} \simeq 0.76\ ,
\end{equation}
where the positive sign has been chosen for the square-root.  The mixing angle
is close to the so-called magic mixing angle $\tan \Theta_V = 1/\sqrt 2 \approx
0.71$, where the $\phi$ meson is a pure $\bar s s$ state.

The measured $\phi \rightarrow \rho \pi$ branching ratio gives $ \left| h_\rho
\right | = 0.05$, where
\begin{equation}\label{eq:h}
h_\rho = {g_1 \over \sqrt 2} \sin \Theta_V - {g_2 \over \sqrt 3} \cos \Theta_V.
\end{equation}
The value of $h_\rho$ alone does not determine the coupling constants $g_1$ and
$g_2$. $h_\rho$ vanishes in the large-$N_c$ limit, so the value of $h_\rho$ is
sensitive to deviations of $\Theta_V$ and $g_1/g_2$ from their large-$N_c$
values. In addition, the chiral loop correction to the $\phi\rho\pi$ coupling
may be significant~\cite{jmw}.  The smallness of $h_\rho$ implies that
$g_1/g_2$ is close to its large-$N_c$ value of $2/\sqrt3$.

Like the decay $D_s^* \rightarrow D_s \pi^0$, at leading order in chiral
perturbation theory the amplitude for the isospin violating decay $\phi
\rightarrow \omega \pi^0$ arises from a tree diagram involving $\eta\pi$
mixing~\cite{CW}. This Feynman diagram gives
\begin{equation}\label{eq:rate}
\Gamma(\phi \rightarrow \omega \pi^0)=\left({m_d-m_u\over m_s}
\right)^2{h_{\eta}^2\over 8\pi}\left({m_{\omega} \over
m_{\phi}}\right){\left|{\bf p}_{\pi}\right|^3 \over f^2}
\end{equation}
where the $\phi\omega\eta$ coupling is
\begin{equation}\label{eq:ha}
h_\eta = {g_2\over2\sqrt3} \sin 2 \Theta_V - {g_1\over \sqrt2} \cos 2 \Theta_V.
\end{equation}
Using~\cite{GL} $(m_d-m_u)/ m_s=1/44$ in Eq.~(\ref{eq:rate}) implies a
branching ratio, ${\rm Br}(\phi\rightarrow\omega\pi)=4.4\times
10^{-6}(h_{\eta}/0.05)^2$. In principle the value of the angle $\Theta_V$ in
Eq.~(\ref{eq:19}), the measured $\phi\rho\pi$ coupling and a measurement of the
branching ratio for $\phi\rightarrow \omega\pi$ would determine
$g_{1,2}$. However corrections from higher orders in chiral perturbation theory
may be important. It is desireable to have an alternate determination of these
couplings.  Some other $\phi$ decays which might have been used to determine
$g_{1,2}$, such as $\phi \rightarrow K^* K$, are not kinematically allowed.

\begin{figure}
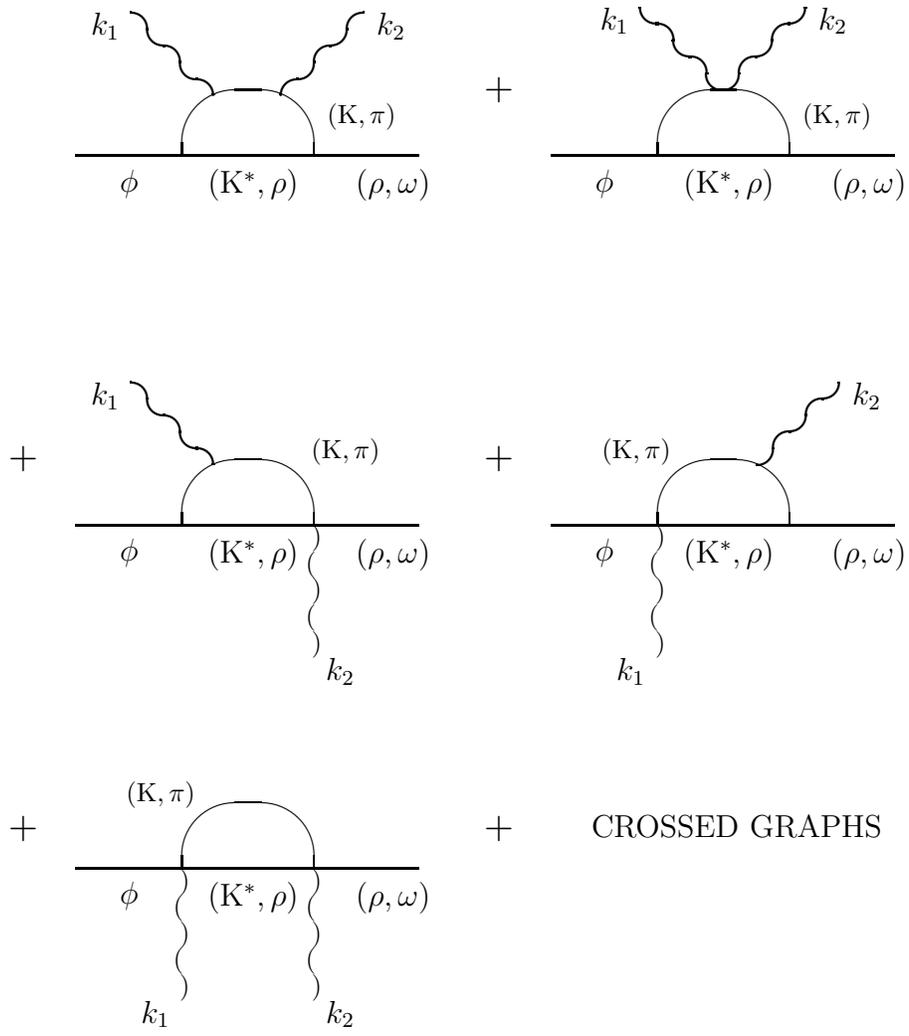

\caption{Loop Graphs for $\phi \rightarrow \rho \gamma \gamma$ and
$\phi \rightarrow \omega \gamma \gamma$ that contribute at leading order in
chiral perturbation theory. Graphs with a photon attached to a
vector meson line vanish in $v\cdot \epsilon=0$ gauge.}
\label{fig:1}
\end{figure}

The coupling constants $g_{1,2}$ can be determined from the radiative $\phi$
decays $\phi \rightarrow \rho \gamma \gamma$, and $\phi \rightarrow \omega
\gamma \gamma$. At leading order in chiral perturbation theory, there are two
contributions; the loop graphs in Fig.~\ref{fig:1} and the anomaly graph of
Fig.~\ref{fig:2}, and there are no counterterms at this order.  The computation
of the loop graphs is straightfoward. We will evaluate the graphs in the limit
that the photon momentum is small compared with the meson mass. In the
large-$N_c$ limit, the non-zero loop graphs have $K$-meson loops, and the
photon momentum is smaller than $M_K$ for the $\phi \rightarrow \rho \gamma
\gamma$ and $\phi \rightarrow \omega \gamma \gamma$ decays.  The decay
amplitude from the loop graph is\footnote{Vector meson states are normalized to
unity. To convert to the usual covariant normalization, multiply
Eq.~(\ref{eq:aloop}) by $\sqrt{4 m_\rho m_\phi}$.}
\begin{eqnarray}\nonumber
A_{\rm loop} &=& -i {\alpha g_2^2 \over 12 f^2}\ {\cal A}\ \epsilon_\phi^\mu
\epsilon^{*\nu}_\rho \Biggl\{ \\ && \left[f^{(1)}_{\mu \lambda}
f^{(2)}_\nu{}^\lambda + f^{(1)}_{\nu \lambda} f^{(2)}_\mu{}^\lambda\right]
\label{eq:aloop} \\ &&+ g_{\mu\nu} \left[- f^{(1)}_{\alpha \beta} f^{(2)\alpha
\beta} + 10 v^\alpha v^\beta f^{(1)}_{\alpha \lambda}
f^{(2)}_\beta{}^\lambda\right] \nonumber \\ && - 3 v^\alpha v^\beta \left[
f^{(1)}_{\alpha \mu} f^{(2)}_{\beta \nu} + f^{(2)}_{\alpha \mu} f^{(1)}_{\beta
\nu}\right]\Biggr\} \nonumber
\end{eqnarray}
where
\begin{equation}
f^{(i)}_{\alpha\beta} = \epsilon_{i\alpha}^* k_{i\beta} - \epsilon_{i\beta}^*
k_{i\alpha},
\end{equation}
and $\epsilon_i$ and $k_i$ are the polarization and momentum of photon $i$.
The constant $\cal A$ is
\begin{equation}
{\cal A}_\rho = \left[{1\over\sqrt 2} {g_1\over g_2} \sin \Theta_V + {1\over 2
\sqrt 3} \cos \Theta_V \right ]{1\over M_K}
\end{equation}
for $\phi \rightarrow \rho \gamma \gamma$, and
\begin{eqnarray}\nonumber
{\cal A}_\omega &=& \left[{1\over\sqrt 6} {g_1\over g_2}\cos 2 \Theta_V -
\left({1\over 12} - {g_1^2\over 2g_2^2} \right) \sin 2\Theta_V \right] {1\over
M_K}\\ &&- \left[\sqrt {2\over 3} {g_1\over g_2}\cos 2 \Theta_V +\left({1\over
3} - {g_1^2\over 2g_2^2} \right) \sin 2\Theta_V \right] {1\over M_\pi}
\end{eqnarray}
for $\phi \rightarrow \omega \gamma \gamma$.  In the large-$N_c$ limit, the
constants have the simpler form
\begin{equation}\label{eq:AlargeN}
A_\rho = A_{\omega} ={1\over \sqrt 2\ M_K}.
\end{equation}

Neglecting isospin nonconservation arising from the up-down quark mass
difference the amplitude from the anomaly graph is
\begin{eqnarray}
&&{\cal A}_{\rm anomaly} = -i h a {2\alpha \over \pi f^2} \epsilon_\phi^\mu
\epsilon^{*\nu}_\rho {1\over 2 k_1 \cdot k_2 - M^2_{\pi,\eta}} \\ &&\times
\epsilon_{\mu \nu \lambda \sigma}\ \epsilon_{\alpha \beta \tau \rho} \
\epsilon_1^{*\alpha}\, \epsilon_2^{*\beta}\, k_1^\tau\, k_2^\rho\ \left( k_1 +
k_2\right)^\lambda\ v^\sigma.
\end{eqnarray}
$M_{\pi,\eta}$ is the pion or eta mass, $h$ is the $\phi \rho \pi$ coupling
$h_\rho$ (Eq.~\ref{eq:h}) or the $\phi \omega \eta $ coupling $h_{\eta}$
(Eq.~\ref{eq:ha}) and $a_{\rho,\eta}=1,1/\sqrt3$ is the anomaly
coefficient. Since the parameters in the chiral Lagrangian are close to their
large-$N_c$ values, we will evaluate the radiative decay rate in this
limit. The anomaly graph and the loop graph are of the same order in the
$1/N_c$ expansion.

The anomaly amplitude is antisymmetric in $\epsilon_\phi \leftrightarrow
\epsilon^*_\rho$, whereas the loop amplitude is symmetric. As a result, there
is no interference term for the spin averaged decay rate, so we will consider
the loop and anomaly decay rates separately. It is convenient to use the
dimensionless variables
\begin{equation}
\hat s = {(k_1+k_2)^2\over m_\phi^2},\ x = {2 (E_1 - E_2)\over m_\phi},\ r =
{m^2_{\rho,\omega}\over m_\phi^2}.
\end{equation}
where $k_1$ and $k_2$ are the four-momenta of the two photons, and $E_1$ and
$E_2$ are their energies in the rest frame of the $\phi$ meson.  The allowed
region in the Dalitz plot in the $x\hat s$ plane is $0\le x \le \left[(1+r-\hat
s)^2-4r\right]^{1/2}$, $0 \le \hat s \le (1-\sqrt r)^2$.

The differential decay distribution from the loop graph is
\begin{eqnarray}\nonumber
{d\Gamma_{\rm loop} \over dx\, d\hat s} &=& {m_\phi^6 m_{\rho,\omega} g_2^4 \pi
\over 864 M_K^2} \left( {\alpha \over 16 \pi^2 f^2}\right)^2\\ \times&& \left[
69 \hat s^2 - {89\over 4}\hat s \left(1-r+\hat s-x\right) \left(1-r+\hat s+x
\right)\right. \\ &&\left. + {99\over 32} \left(1-r+\hat s-x\right)^2
\left(1-r+\hat s+x \right)^2 \right], \nonumber
\end{eqnarray}
where the large-$N_c$ values Eq.~(\ref{eq:AlargeN}) have been used.  The
differential decay distribution from the anomaly graph is
\begin{eqnarray}
{d\Gamma_{\rm anomaly} \over dx\, d\hat s} &=& {2m_\phi^8 m_{\rho,\omega} h^2
a^2\hat s^2\over 3 \pi} \left( {\alpha \over 16 \pi^2 f^2}\right)^2 \nonumber\\
&&\times\ \ {\left(1+r-\hat s\right)^2 - 4 r \over \left( m_\phi^2 \hat s -
M^2_{\pi,\eta}\right)^2} .
\end{eqnarray}

Integrating over $x$ from 0 to $\left[(1+r-\hat s)^2-4 r\right]^{1/2}$ gives
\begin{eqnarray}\nonumber
&&{d\Gamma_{\rm loop} \over d\hat s} = {m_\phi^6 m_{\rho,\omega} g_2^4 \pi
\over 51840 M_K^2} \left( {\alpha \over 16 \pi^2 f^2}\right)^2\nonumber \\
&&\quad\times\quad \left[(1+r-\hat s)^2 - 4 r \right]^{1/2} \\
&&\quad\times\quad \Biggl\{ 99 - 396\,r + 594\,{r^2} - 396\,{r^3} +
99\,{r^4}\nonumber\\ && - 296\,\hat s + 196\,r\,\hat s + 496\,{r^2}\, \hat s -
396\,{r^3}\, \hat s + 2164\,{\hat s^2}\nonumber\\ && + 196\,r\,{\hat s^2} +
594\,{r^2}\,{\hat s^2} - 296\,{\hat s^3} - 396\,r\,{\hat s^3} + 99\,{\hat s^4}
\Biggr\},\nonumber
\end{eqnarray}
\begin{eqnarray}
{d\Gamma_{\rm anomaly} \over d\hat s} &=& {2 m_\phi^8 m_{\rho,\omega} h^2
a^2\hat s^2\over 3 \pi} \left( {\alpha \over 16 \pi^2 f^2}\right)^2 \nonumber\\
&&\times\ \ {\left[\left(1+r-\hat s\right)^2 - 4 r\right]^{3/2} \over \left(
m_\phi^2 \hat s - M^2_{\pi,\eta}\right)^2} .
\end{eqnarray}

Total decay rates are obtained by integrating $d\Gamma/d\hat s$ from $\hat s=0$
to $\hat s=(1-\sqrt r)^2$.  The branching ratios from the loop graph for $\phi
\rightarrow \rho \gamma \gamma$ and $\phi \rightarrow \omega \gamma \gamma$ are
$5.8\times 10^{-9} (g_2/0.75)^4$ and $4.2\times 10^{-9} (g_2/0.75)^4$,
respectively. The non-relativistic quark model predicts that $g_2=1$, and the
somewhat smaller value $g_2\approx 0.75$ is suggested by the chiral quark
model~\cite{MG}. These branching ratios are quite small, but the radiative
decays should be observable at the $\phi$ factory.

The anomaly graph for $\phi \rightarrow \omega \gamma \gamma$ involves the
unknown coupling $h_\eta$, which vanishes in the large-$N_c$ limit. Since
$h_\eta$ and $h_\rho$ both vanish in the large-$N_c$ limit we will use
$h_\eta\approx h_\rho$ to estimate the importance of the anomaly graph relative
to the loop graph. With this assumption for $h_\eta$, the anomaly graph is much
smaller than the loop graph over the entire Dalitz plot, and can be neglected
(The ratio of decay widths is $\Gamma_{\rm anomaly}/\Gamma_{\rm loop} = 5\times
10^{-4} (h_\eta/0.05)^2 (0.75/g_2)^4$.) Thus a measurement of the $\phi
\rightarrow \omega \gamma \gamma$ decay rate\footnote{ For $\hat s$ very near
$M_\pi^2/m_\phi^2 = 0.0175$ the $\phi \rightarrow\omega\gamma\gamma$ rate is
dominated by the isospin violating process
$\phi\rightarrow\omega\pi\rightarrow\omega\gamma\gamma$ which is expected to
have a branching ratio around $4\times 10^{-6}$} allows one to determine the
coupling constant $g_2$, and $g_1 = 2 g_2/\sqrt3$.  It might even be possible
to compare the chiral peturbation theory prediction for the differential decay
distribution with experiment if $g_2$ is reasonably large, and the $\phi$
factory achieves its design luminosity.

The situation is quite different for $\phi \rightarrow \rho \gamma \gamma$.
The branching ratio from the anomaly graph for $\phi \rightarrow \rho \gamma
\gamma$ is the same as the branching ratio of 4.3\% for the process $\phi
\rightarrow \rho^0 \pi^0$, since the $\pi^0$ can be on-shell, and it decays
almost exclusively into two photons.  Thus the total decay rate from the
anomaly overwhelms that from the loop diagram. However the rate from the
anomaly graph is concentrated in a band in the $x\hat s$ plane centered around
$\hat s = M_\pi^2/m_\phi^2 = 0.0175$. The ratio of the loop to the total decay
distribution is show in Figs.~(\ref{fig:4},\ref{fig:5}). There are regions in
the Dalitz plot where the loop graph is the dominant decay mechanism. Measuring
the radiative decay rate in this region should allow one to determine
$\Gamma_{\rm loop}$, and thus the coupling constant $g_2$.

The rate for $\phi \rightarrow \rho \gamma \gamma$ near the pion-pole is mainly
from the anomaly graph.  A measurement of the rate in this region of the Dalitz
plot probes the $q^2$ dependence of the $\pi^0\rightarrow \gamma\gamma$
amplitude, which is theoretically interesting. The $q^2$ dependence arises from
higher order terms in the meson chiral Lagrangian that contain an
$\epsilon$-symbol, and also from loop graphs involving the Wess-Zumino term.

This work was supported in part by the Department of Energy under Grant
Nos.~DOE-FG03-90ER40546 and DE-FG03-92-ER40701. A.M.\ was supported in part by
the PYI program, through Grant No.~PHY-8958081 from the National Science
Foundation.

\onecolumn
\widetext

\input FEYNMAN

\textheight 800pt \textwidth 530pt

\moveright1.75in\hbox{ 
\begin{picture}(10000,10000)
\THICKLINES


\drawline\fermion[\E\REG](-3000,8000)[4000]
\global\Xeight=\pfrontx
\global\Yeight=\pfronty
\global\Xseven=\pfrontx
\global\Yseven=\pfronty
\global\Xsix=\pfrontx
\global\Ysix=\pfronty
\global\Xone=\pbackx
\global\Yone=\pbacky
\global\Xtwo=\pfrontx
\global\Ytwo=\pfronty
\global\advance\Xone by 2500
\global\advance\Xtwo by 2000
\global\advance\Ytwo by -500
\put(\Xtwo,\Ytwo) {\makebox(0,0)[t]{\large $\phi$}}

\thinlines
\put(\Xone,\Yone){\oval (5000,5000)[t]}
\THICKLINES

\drawline\fermion[\E\REG](\pbackx,\pbacky)[5000]
\global\advance\Xtwo by 4700
\put(\Xtwo,\Ytwo) {\makebox(0,0)[t]{\large $({\rm K}^*,\rho)$}}

\drawline\fermion[\E\REG](\pbackx,\pbacky)[4000]
\global\advance\Xtwo by 5300
\put(\Xtwo,\Ytwo) {\makebox(0,0)[t]{\large $(\rho,\omega)$}}

\global\advance\Xtwo by -1200
\global\advance\Ytwo by 2500
\put(\Xtwo,\Ytwo) {\makebox(0,0)[t]{$({\rm K},\pi)$}}

\global\advance\Xone by -1280
\global\advance\Yone by 2300
\drawline\photon[\NW\FLIPPED](\Xone,\Yone)[5]
\global\Xtwo=\pbackx
\global\Ytwo=\pbacky
\global\advance\Xtwo by -1000
\put(\Xtwo,\Ytwo) {\makebox(0,0)[t]{\large $k_1$}}

\global\advance\Xone by 2560
\drawline\photon[\NE\REG](\Xone,\Yone)[5]
\global\Xtwo=\pbackx
\global\Ytwo=\pbacky
\global\advance\Xtwo by 1000
\put(\Xtwo,\Ytwo) {\makebox(0,0)[t]{\large $k_2$}}

\global\advance\Xseven by 16000
\global\advance\Yseven by 3000
\put(\Xseven,\Yseven) {\makebox(0,0)[t]{\Large $+$}}


\global\Xone=\Xeight
\global\Yone=\Yeight
\global\advance\Xone by 18000

\drawline\fermion[\E\REG](\Xone,\Yone)[4000]
\global\Xone=\pbackx
\global\Yone=\pbacky
\global\Xtwo=\pfrontx
\global\Ytwo=\pfronty
\global\advance\Xone by 2500
\global\advance\Xtwo by 2000
\global\advance\Ytwo by -500
\put(\Xtwo,\Ytwo) {\makebox(0,0)[t]{\large $\phi$}}

\thinlines
\put(\Xone,\Yone){\oval (5000,5000)[t]}
\THICKLINES
\global\Xthree=\Xone
\global\Ythree=\Yone

\drawline\fermion[\E\REG](\pbackx,\pbacky)[5000]
\global\advance\Xtwo by 4700
\put(\Xtwo,\Ytwo) {\makebox(0,0)[t]{\large $({\rm K}^*,\rho)$}}

\drawline\fermion[\E\REG](\pbackx,\pbacky)[4000]
\global\advance\Xtwo by 5300
\put(\Xtwo,\Ytwo) {\makebox(0,0)[t]{\large $(\rho,\omega)$}}

\global\advance\Xtwo by -1200
\global\advance\Ytwo by 2500
\put(\Xtwo,\Ytwo) {\makebox(0,0)[t]{$({\rm K},\pi)$}}

\global\advance\Ythree by 2500
\drawline\photon[\NW\REG](\Xthree,\Ythree)[5]
\global\Xtwo=\pbackx
\global\Ytwo=\pbacky
\global\advance\Xtwo by -1000
\put(\Xtwo,\Ytwo) {\makebox(0,0)[t]{\large $k_1$}}

\drawline\photon[\NE\FLIPPED](\Xthree,\Ythree)[5]
\global\Xtwo=\pbackx
\global\Ytwo=\pbacky
\global\advance\Xtwo by 1000
\put(\Xtwo,\Ytwo) {\makebox(0,0)[t]{\large $k_2$}}


\global\advance\Xsix by -2000
\global\advance\Ysix by -11000
\put(\Xsix,\Ysix) {\makebox(0,0)[t]{\Large $+$}}

\global\Xone=\Xeight
\global\Yone=\Yeight
\global\advance\Yone by -14000
\drawline\fermion[\E\REG](\Xone,\Yone)[4000]
\global\Xone=\pbackx
\global\Yone=\pbacky
\global\Xtwo=\pfrontx
\global\Ytwo=\pfronty
\global\advance\Xone by 2500
\global\advance\Xtwo by 2000
\global\advance\Ytwo by -500
\put(\Xtwo,\Ytwo) {\makebox(0,0)[t]{\large $\phi$}}

\thinlines
\put(\Xone,\Yone){\oval (5000,5000)[t]}
\THICKLINES

\drawline\fermion[\E\REG](\pbackx,\pbacky)[5000]
\global\advance\Xtwo by 4700
\put(\Xtwo,\Ytwo) {\makebox(0,0)[t]{\large $({\rm K}^*,\rho)$}}

\global\Xfour=\pbackx
\global\Yfour=\pbacky
\drawline\photon[\S\REG](\pbackx,\pbacky)[5]
\global\Xthree=\pbackx
\global\Ythree=\pbacky
\global\advance\Xthree by 1000
\put(\Xthree,\Ythree) {\makebox(0,0)[t]{\large $k_2$}}

\drawline\fermion[\E\REG](\Xfour,\Yfour)[4000]
\global\advance\Xtwo by 5300
\put(\Xtwo,\Ytwo) {\makebox(0,0)[t]{\large $(\rho,\omega)$}}

\global\advance\Xtwo by -1800
\global\advance\Ytwo by 3800
\put(\Xtwo,\Ytwo) {\makebox(0,0)[t]{$({\rm K},\pi)$}}

\global\advance\Xone by -1280
\global\advance\Yone by 2300
\drawline\photon[\NW\FLIPPED](\Xone,\Yone)[5]
\global\Xtwo=\pbackx
\global\Ytwo=\pbacky
\global\advance\Xtwo by -1000
\put(\Xtwo,\Ytwo) {\makebox(0,0)[t]{\large $k_1$}}

\global\advance\Yseven by -14000
\put(\Xseven,\Yseven) {\makebox(0,0)[t]{\Large $+$}}


\global\Xone=\Xeight
\global\Yone=\Yeight
\global\advance\Xone by 18000
\global\advance\Yone by -14000
\drawline\fermion[\E\REG](\Xone,\Yone)[4000]
\global\Xone=\pbackx
\global\Yone=\pbacky
\global\Xtwo=\pfrontx
\global\Ytwo=\pfronty
\global\advance\Xone by 2500
\global\advance\Xtwo by 2000
\global\advance\Ytwo by -500
\put(\Xtwo,\Ytwo) {\makebox(0,0)[t]{\large $\phi$}}

\thinlines
\put(\Xone,\Yone){\oval (5000,5000)[t]}
\THICKLINES

\drawline\fermion[\E\REG](\pbackx,\pbacky)[5000]
\global\advance\Xtwo by 4700
\put(\Xtwo,\Ytwo) {\makebox(0,0)[t]{\large $({\rm K}^*,\rho)$}}

\global\Xfour=\pbackx
\global\Yfour=\pbacky
\drawline\photon[\S\REG](\pfrontx,\pfronty)[5]
\global\Xthree=\pbackx
\global\Ythree=\pbacky
\global\advance\Xthree by -1000
\put(\Xthree,\Ythree) {\makebox(0,0)[t]{\large $k_1$}}

\drawline\fermion[\E\REG](\Xfour,\Yfour)[4000]
\global\advance\Xtwo by 5300
\put(\Xtwo,\Ytwo) {\makebox(0,0)[t]{\large $(\rho,\omega)$}}

\global\advance\Xtwo by -8800
\global\advance\Ytwo by 3800
\put(\Xtwo,\Ytwo) {\makebox(0,0)[t]{$({\rm K},\pi)$}}

\global\advance\Xone by 1280
\global\advance\Yone by 2300
\drawline\photon[\NE\FLIPPED](\Xone,\Yone)[5]
\global\Xtwo=\pbackx
\global\Ytwo=\pbacky
\global\advance\Xtwo by 1000
\put(\Xtwo,\Ytwo) {\makebox(0,0)[t]{\large $k_2$}}


\global\advance\Ysix by -14000
\put(\Xsix,\Ysix) {\makebox(0,0)[t]{\Large $+$}}

\global\Xone=\Xeight
\global\Yone=\Yeight
\global\advance\Yone by -27000
\drawline\fermion[\E\REG](\Xone,\Yone)[4000]
\global\Xone=\pbackx
\global\Yone=\pbacky
\global\Xtwo=\pfrontx
\global\Ytwo=\pfronty
\global\advance\Xone by 2500
\global\advance\Xtwo by 2000
\global\advance\Ytwo by -500
\put(\Xtwo,\Ytwo) {\makebox(0,0)[t]{\large $\phi$}}

\thinlines
\put(\Xone,\Yone){\oval (5000,5000)[t]}
\THICKLINES

\drawline\fermion[\E\REG](\pbackx,\pbacky)[5000]
\global\advance\Xtwo by 4700
\put(\Xtwo,\Ytwo) {\makebox(0,0)[t]{\large $({\rm K}^*,\rho)$}}

\global\Xfour=\pbackx
\global\Yfour=\pbacky
\drawline\photon[\S\REG](\pfrontx,\pfronty)[5]
\global\Xthree=\pbackx
\global\Ythree=\pbacky
\global\advance\Xthree by -1000
\put(\Xthree,\Ythree) {\makebox(0,0)[t]{\large $k_1$}}

\drawline\fermion[\E\REG](\Xfour,\Yfour)[4000]
\global\advance\Xtwo by 5300
\put(\Xtwo,\Ytwo) {\makebox(0,0)[t]{\large $(\rho,\omega)$}}

\drawline\photon[\S\REG](\pfrontx,\pfronty)[5]
\global\Xthree=\pbackx
\global\Ythree=\pbacky
\global\advance\Xthree by 1000
\put(\Xthree,\Ythree) {\makebox(0,0)[t]{\large $k_2$}}

\global\advance\Xtwo by -8800
\global\advance\Ytwo by 3800
\put(\Xtwo,\Ytwo) {\makebox(0,0)[t]{$({\rm K},\pi)$}}

\global\advance\Yseven by -14000
\put(\Xseven,\Yseven) {\makebox(0,0)[t]{\Large $+$}}


\global\Xone=\Xeight
\global\Yone=\Yeight
\global\advance\Yone by -25000
\global\advance\Xone by 25000

\put(\Xone,\Yone) {\makebox(0,0)[t]{\large CROSSED GRAPHS}}

\end{picture}

} 

\vskip 4 in
\vskip1in

\textheight 800pt \textwidth 530pt

\moveright2.25in\hbox{ 
\begin{picture}(10000,10000)
\THICKLINES

\drawline\fermion[\E\REG](1000,3000)[6000]
\global\Xone=\pfrontx
\global\Yone=\pfronty
\global\advance\Xone by 3000
\global\advance\Yone by -500
\put(\Xone,\Yone) {\makebox(0,0)[t]{\large $\phi_\mu$}}

\drawline\fermion[\E\REG](\pbackx,\pbacky)[8200]
\global\advance\Xone by 7200
\put(\Xone,\Yone) {\makebox(0,0)[t]{\large $(\rho,\omega)_\nu$}}

\thinlines
\drawline\fermion[\NE\REG](\pfrontx,\pfronty)[5000]
\THICKLINES
\global\Xone=\pmidx
\global\Yone=\pmidy
\global\advance\Xone by -2500
\global\advance\Yone by 500
\put(\Xone,\Yone) {\makebox(0,0){\large $(\pi,\eta)$}}

\drawline\photon[\N\REG](\pbackx,\pbacky)[4]
\global\Xone=\pbackx
\global\Yone=\pbacky
\global\advance\Yone by 1000
\put(\Xone,\Yone) {\makebox(0,0){\large $k_1$}}

\drawline\photon[\E\FLIPPED](\pfrontx,\pfronty)[4]
\global\Xone=\pbackx
\global\Yone=\pbacky
\global\advance\Xone by 1000
\put(\Xone,\Yone) {\makebox(0,0){\large $k_2$}}

\end{picture}
} 

\begin{figure}
\caption{The anomaly graph for $\phi \rightarrow \rho \gamma \gamma$ and
$\phi \rightarrow \omega \gamma \gamma$.}
\label{fig:2}
\end{figure}

\moveright1in\hbox{
\epsffile{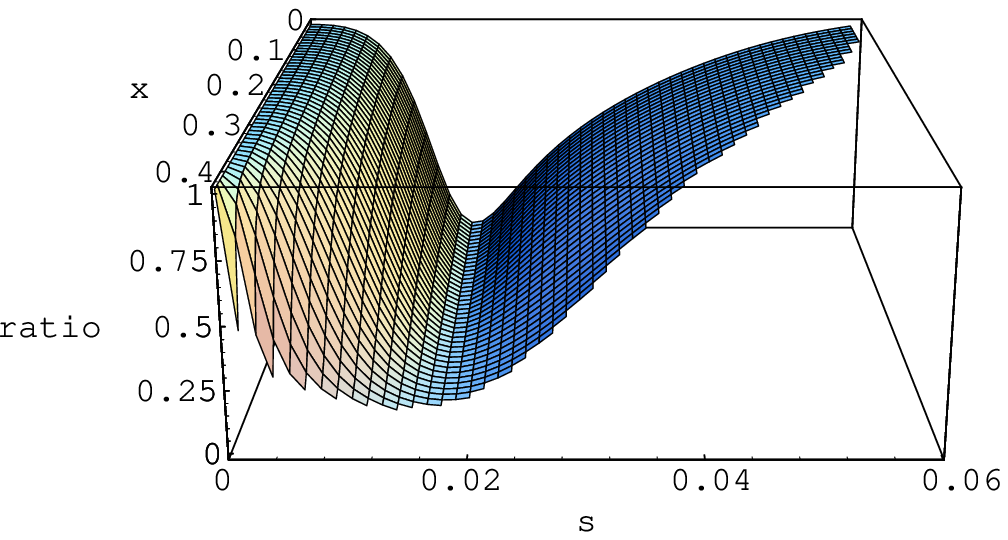}}
\vskip-1in
\begin{figure}
\caption{\hfil
The ratio $d\Gamma_{\rm loop}/dx\,d\hat s/(d\Gamma_{\rm loop}/dx\,d\hat s
+d\Gamma_{\rm anomaly}/dx\,d\hat s)$ over the Dalitz plot. The pion pole is at
$\hat s =0.0175$.}
\label{fig:4}
\end{figure}

\moveright1in\hbox{
\epsffile{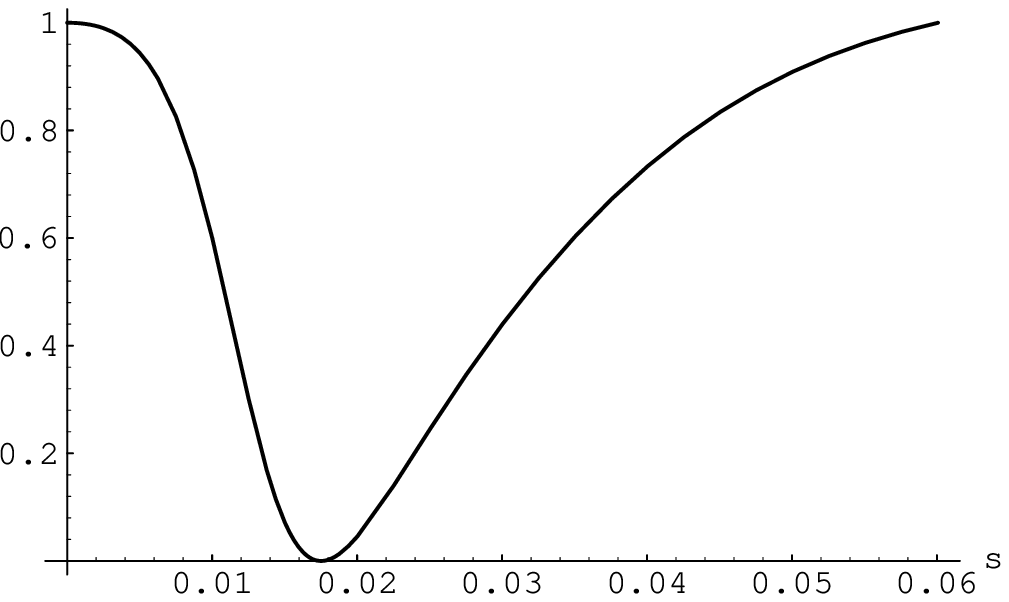}}
\vskip-1in
\begin{figure}
\caption{\hfil
The ratio $d\Gamma_{\rm loop}/d\hat s/(d\Gamma_{\rm loop}/d\hat s
+d\Gamma_{\rm anomaly}/d\hat s)$ as a function of $\hat s$.
The pion pole is at $\hat s =0.0175$.}
\label{fig:5}
\end{figure}

\end{document}